\documentclass[conference]{IEEEtran}
\IEEEoverridecommandlockouts
\usepackage{cite}
\usepackage{amsmath,amssymb,amsfonts}
\usepackage{algorithmic}
\usepackage{graphicx}
\usepackage{float}
\usepackage{subfigure}
\usepackage{textcomp}
\usepackage{xcolor}
\def\BibTeX{{\rm B\kern-.05em{\sc i\kern-.025em b}\kern-.08em
    T\kern-.1667em\lower.7ex\hbox{E}\kern-.125emX}}
\begin{document}

\title{Row-wise Accelerator for Vision Transformer}
\bstctlcite{IEEEexample:BSTcontrol}

\author{\IEEEauthorblockN{Hong-Yi Wang, and Tian-Sheuan Chang, \textit{Senior Member, IEEE}}
\IEEEauthorblockA{\textit{Institute of Electronics, National Yang Ming Chiao Tung University,}
Hsinchu, Taiwan \\
}
\thanks{H.-Y. Wang, and T.-S. Chang, "Row-wise accelerator for vision transformer", IEEE AICAS, 2022}
}
\maketitle

\begin{abstract}
Following the success of the natural language processing, the transformer for vision applications has attracted significant attention in recent years due to its excellent performance. However, existing deep learning hardware accelerators for vision cannot execute this structure efficiently due to significant model architecture differences. As a result, this paper proposes the hardware accelerator for vision transformers with row-wise scheduling, which decomposes major operations in vision transformers as a single dot product primitive for a unified and efficient execution. Furthermore, by sharing weights in columns, we can reuse the data and reduce the usage of memory. The implementation with TSMC 40nm CMOS technology only requires 262K gate count and 149KB SRAM buffer for 403.2 GOPS throughput at 600MHz clock frequency.

\end{abstract}


\section{Introduction}
Deep learning in computer vision has long been dominated by convolutional neural networks (CNNs) based backbone\cite{krizhevsky2012imagenet} for its revolutionary performance. On the other hand, natural language processing (NLP) use Transformer~\cite{vaswani2017attention} that no longer relies on convolution but self-attention to model long-term data dependencies. Its enormous success makes it a viable competitor in computer vision tasks as well. As a result, various vision transformer models\cite{dosovitskiy2020image,liu2021swin, han2021transformer,wang2021pyramid} have been developed that show competitive or even better performance than current CNN based models. However, real-time executions of these models suffer from high computational complexity and memory bandwidth, which demands hardware acceleration. However, existing deep learning hardware accelerators\cite{chen2016eyeriss, chen2019eyeriss, biswas2018conv} are optimized for CNN based models, which is not suitable for vision transformer models due to the significant difference in the favored architecture structure and parameters. Directly executing vision transformer models on these accelerators will result in low hardware utilization.

To address the above issue, this paper proposes a hardware accelerator for vision transformers. To tailor the hardware design to vision transformers, we first analyze the common features of different models and compare them with existing CNN models. Based on this analysis, we propose a row-wise scheduling that uses the dot product as a primitive for major operations in the model to attain high hardware utilization. Besides, the proposed hardware broadcast weight for computations that can share weights for higher data reuse and smaller buffer size. The implementation on TSMC 40nm CMOS process can achieve 403.2GOPS throughput when running at 600Mhz clock frequency.

The remainder of the paper is structured as follows. Section II goes over the associated tasks. Section III examines the common characteristics of several vision transformer models. The proposed design is shown in Section IV. Section V presents the experimental data as well as comparisons with other studies. Finally, Section VI brings this paper to a close.

\section{Related Work}

\subsection{Transformer}

\begin{figure}[htb]
\centering
\includegraphics[width=0.48\textwidth]{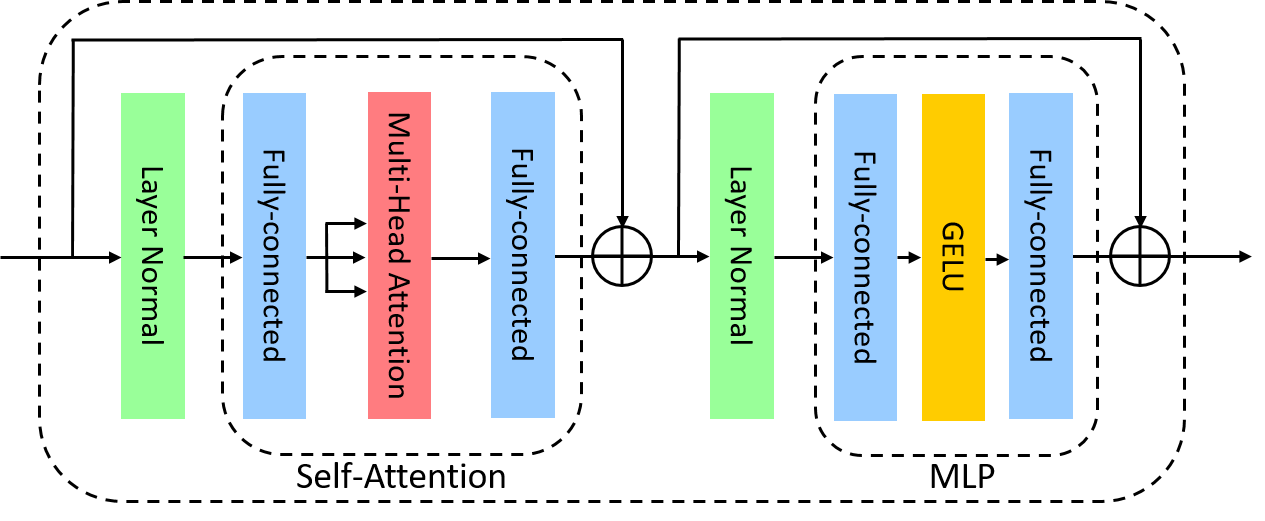}
\caption{Transformer encoder architecture}
\label{transformer}
\end{figure}

Transformer adopts the encoder and decoder architecture. The encoder consists of multiple encoding layers to generate the relationship between the inputs, while the decoder consists of multiple decoding layers that do the opposite. Fig.~\ref{transformer} shows an example of one layer of encoder. In which, an encoding layer consists of three kinds of blocks, self-attention, feed forward and normalization.The self-attention block will figure out the relationship between different inputs to generate output encodings as shown in (\ref{eq self attention}). 
\begin{equation}
\label{eq self attention}
    Attention(Q,K,V) = SoftMax(QK^{T}/\sqrt d_{k})V
\end{equation}
Then, each output encoding is further processed by the feed forward block that uses two successive fully connected layers with Gaussian error linear units (GELU) \cite{hendrycks2016gaussian} as its activation function. Its output is normalized by the normalization block that uses layer normalization. Besides, they would use the skip connection in every block. 

The fully connected layers in the vision transformer share their weight along the spatial dimension and apply fully connection on the channel dimension. Thus, these fully connected layers are the 1×1 convolutions. However, to keep naming consistency with other papers, the name of the fully connected layer is still used below.

\subsection{Vision Transformer}
With Transformer's success in nature language processing, Transformer has also been introduced to the vision application~\cite{dosovitskiy2020image}, which gains competitive or better performance than previous convolutional neural network based models.
For the classification task on ImageNet, lots of vision transformers such as ViT~\cite{dosovitskiy2020image}, TNT~\cite{han2021transformer}, Swin~\cite{liu2021swin} and PVT~\cite{wang2021pyramid}  use the encoder of Transformer. These models differ on the detailed structure of the encoder layer.  %

\subsection{Hardware Accelerators for CNNs and Transformer in NLP}
Various hardware accelerators have been proposed for CNN model execution. They exploit the high parallelism of CNN computation with hundreds of PEs for massive computation, and reuse input/weights/activation in different ways to reduce the external memory bandwidth requirement\cite{chen2014dadiannao,chen2014diannao,chen2016eyeriss,chen2019eyeriss}. The PEs are arranged in a 1D or 2D array, optimized for widely used 3×3 convolutions and reconfigured for other convolution types if needed but with lower hardware utilization. In contrast, 3×3 convolutions are seldom used in vision transformers. In vision transformers, their main structures don’t involve CNN. They only use CNN in the beginning as a kind of feature extraction method for image pre-processing. Besides, the kernel size is 4×4 in vision transformers. Thus, if directly executing these models on these accelerators, their  hardware utilization would be lower than 66\%. Not to mention that these designs do not support multi-head attention.

Several works \cite{lu2020hardware,li2020ftrans,ham20203} have also proposed for transformer accelerator in NLP. No data reuse would make it perform poorly in vision. Additionally, the configuration of parameters in NLP (\emph{ sequence length $\times$ channel }) is different from that in vision (\emph{ height $\times$ width $\times$ channel }) .  Thus, existing designs are not suitable for vision transformers.

\section{Model Analysis for Vision Transformers}

\subsection{Common Features in Vision Transformer}
Table~\ref{table:compared2} shows the common configuration parameters for different vision transformer models. First, since these models are for ImageNet classification, the input sizes for different layers are a multiple of 7 due to the 224×224 image size in ImageNet. Besides, their channel numbers were almost always a multiple of 96 except for PVT, which is a multiple of 64. These configuration numbers are quite different from CNN based classification models.

For the model structure, their backbones are the encoder of the transformer, which composes of multi-head attention blocks, residual connection, and fully connected layers. In addition, compared with the traditional CNN model as shown in Table~\ref{table:compared}, activation function of vision transformers is GELU instead of ReLU. Moreover, their normalization functions were layer normalization instead of batch normalization. 

Based on these analysis results, the hardware design will optimize for these common configurations and structures.

\begin{table}[htb]
\caption{Common features in vision transformer models}
\label{table:compared2}
\centering
\begin{tabular}{|l|l|c|c|c|}
\hline
     & features                                                                     & \multicolumn{1}{l|}{channels} & \multicolumn{1}{l|}{input size} & \multicolumn{1}{l|}{conv size} \\ \hline
ViT  &  first vision transformer                  & 768                           & 14                              & 16                                    \\ \hline
TNT  & extra patches transformer                                                      & 768                           & 14                              & 16                                    \\ \hline
Swin & \begin{tabular}[c]{@{}l@{}}windows transformer \\ shifted windows\end{tabular} & 96×                           & 7×                              & 4                                     \\ \hline
PVT  & image pyramid                                                                  & 64×                           & 14                              & 4×                                    \\ \hline
\end{tabular}
\end{table}

\begin{table}[htb]
\caption{Comparison between vision transformer and CNN based model}
\centering
\label{table:compared}
\begin{tabular}{|l|c|c|}
\hline
                     & Vision Transformer  & CNN    \\ \hline
Activation function  & GELU                    & ReLU                       \\ \hline
Normalization        & layer normal            & batch normal               \\ \hline
Main calculation     & fully-connected         & convolution                \\ \hline
Similarity           & \multicolumn{2}{c|}{short cut (residual)}            \\ \hline
\end{tabular}
\end{table}

\begin{figure}[htb]
\centering
\includegraphics[width=0.4\textwidth]{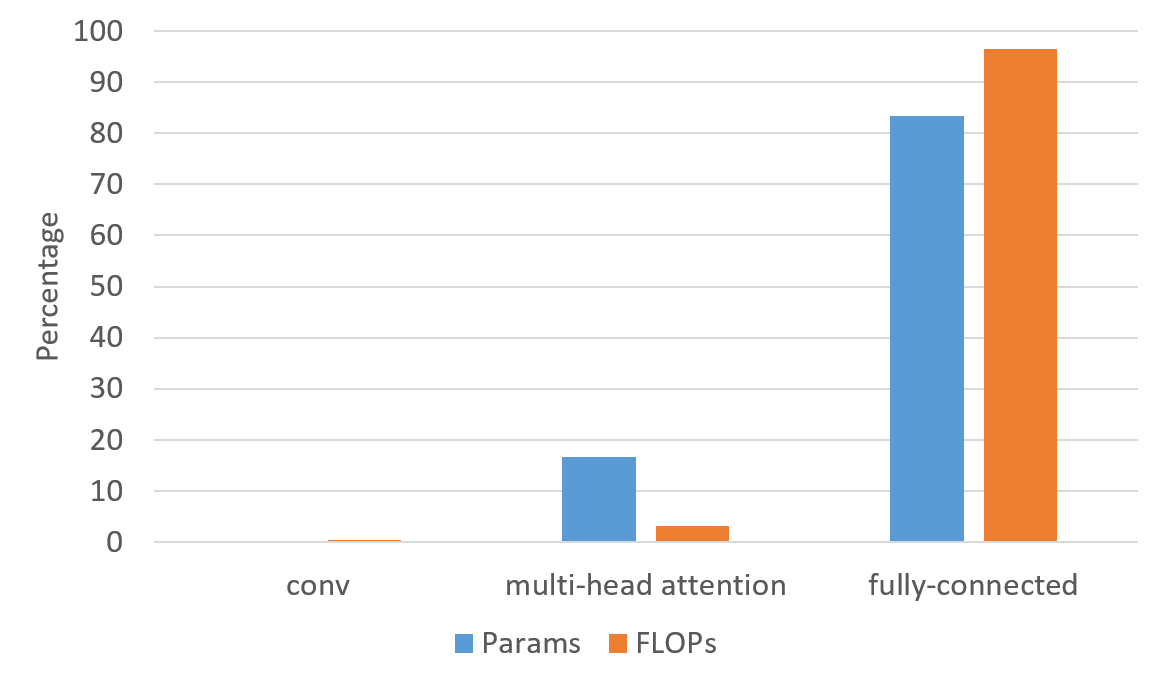}
\caption{Distributions of parameters and FLOPs in Swin-T for different layers.}
\label{distribution layers}
\end{figure}

\subsection{Analysis of FLOPs and Parameters Distribution in Vision Transformers}

To facilitate the analysis, we decompose the model, Swin-T, into three layers according to their actual functions, including the convolution layer, fully connected layer, and multi-head attention layer. Fig.~\ref{distribution layers} shows the distribution analysis of FLOPS and parameters for different layer types .%

From the viewpoint of layer types as shown in Fig.~\ref{distribution layers}, more than 97\% of the FLOPs and more than 83\% of number of parameters are occupied by the fully connected layer. Both the convolution layer and multi-head attention layer only occupy a very small fraction. Thus, the design of computation unit and buffer size should optimize for the fully connected layer.

\section{Proposed Architecture}

\subsection{Row-wise Scheduling}
\begin{figure}[htb]
\centering
\includegraphics[width=0.44\textwidth]{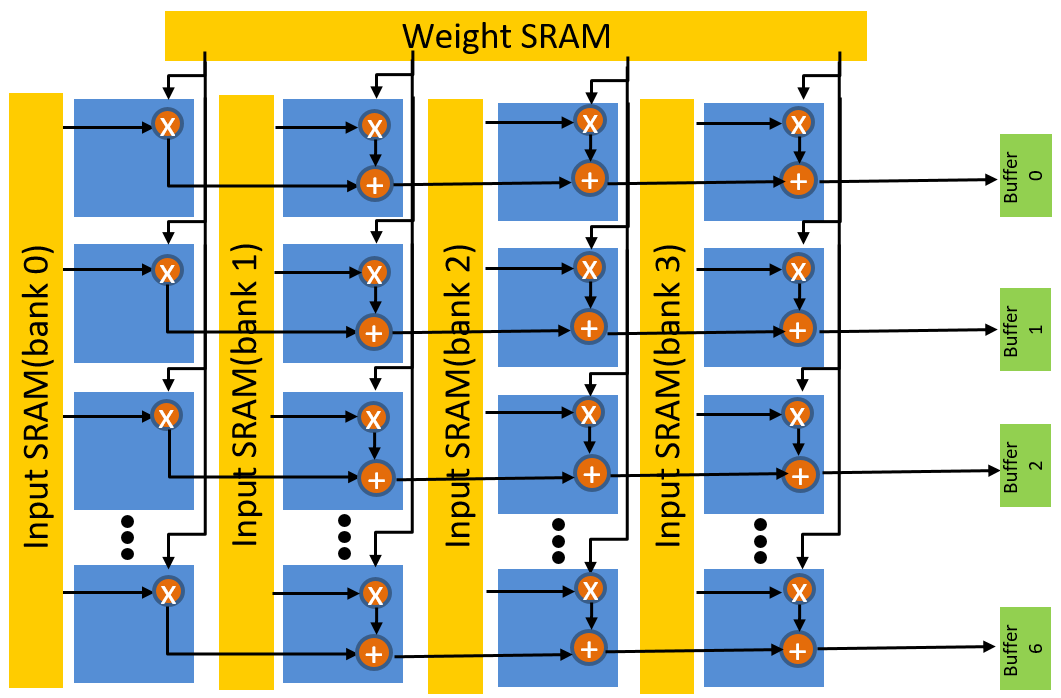}
\caption{A PE block}
\label{PE}
\end{figure}

For the three major layers, multi-head attention and convolution are matrix multiplication, while the fully connected layer is a matrix vector product. To unify these computation types and fit different configuration parameters, we decompose all these operations as a more basic dot product to build our processing element (PE) block as shown in Fig.~\ref{PE}.

A PE block contains 7 PE rows in a block and each row has 4 multiplier-and-accumulator (MAC) units. In this paper, we use 12 PE blocks for model execution. These numbers are selected to fit the common configuration parameters as shown before. Thus, one row can compute a dot product on 2 vectors with both sizes of 4, which can fit different operation requirements due to the decomposition. This also enables a very high hardware utilization and simple dataflow to optimize the overall computing time. 

In a PE block, each weight data are broadcast to all multipliers from top to bottom to share weight. Each MAC will receive different inputs from the input SRAM to support the fully connected layer. The multiplication results are accumulated in a horizontal direction and saved to the local buffer for later more accumulation. 

\subsection{Overall Architecture}
\begin{figure}[htb]
\centering
\includegraphics[width=0.48\textwidth]{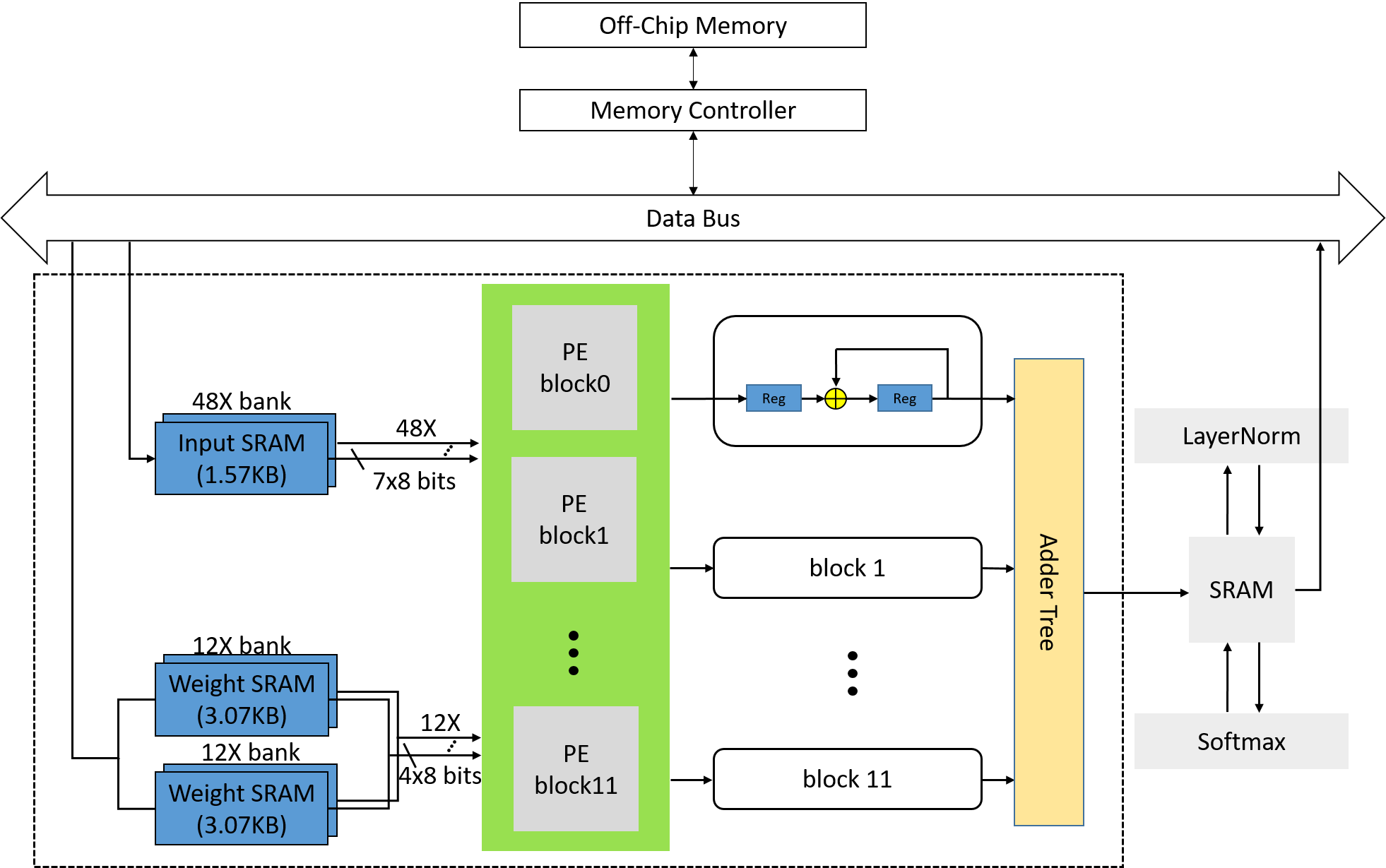}
\caption{Proposed overall architecture for vision transformer}
\label{main}
\end{figure}

Fig.~\ref{main} shows the proposed overall architecture. The result from each PE block is accumulated in an accumulator block and further summed together with the adder tree for the large number of input channel cases. Then the result will be post-processing by layer normalization or softmax depending on the layer types. The final result will be sent to off-chip memory.

\subsection{Convolution Layer}
\begin{figure}[htb]
\centering
\subfigure[]{
\label{PE for convolution}
\includegraphics[width=0.23\textwidth]{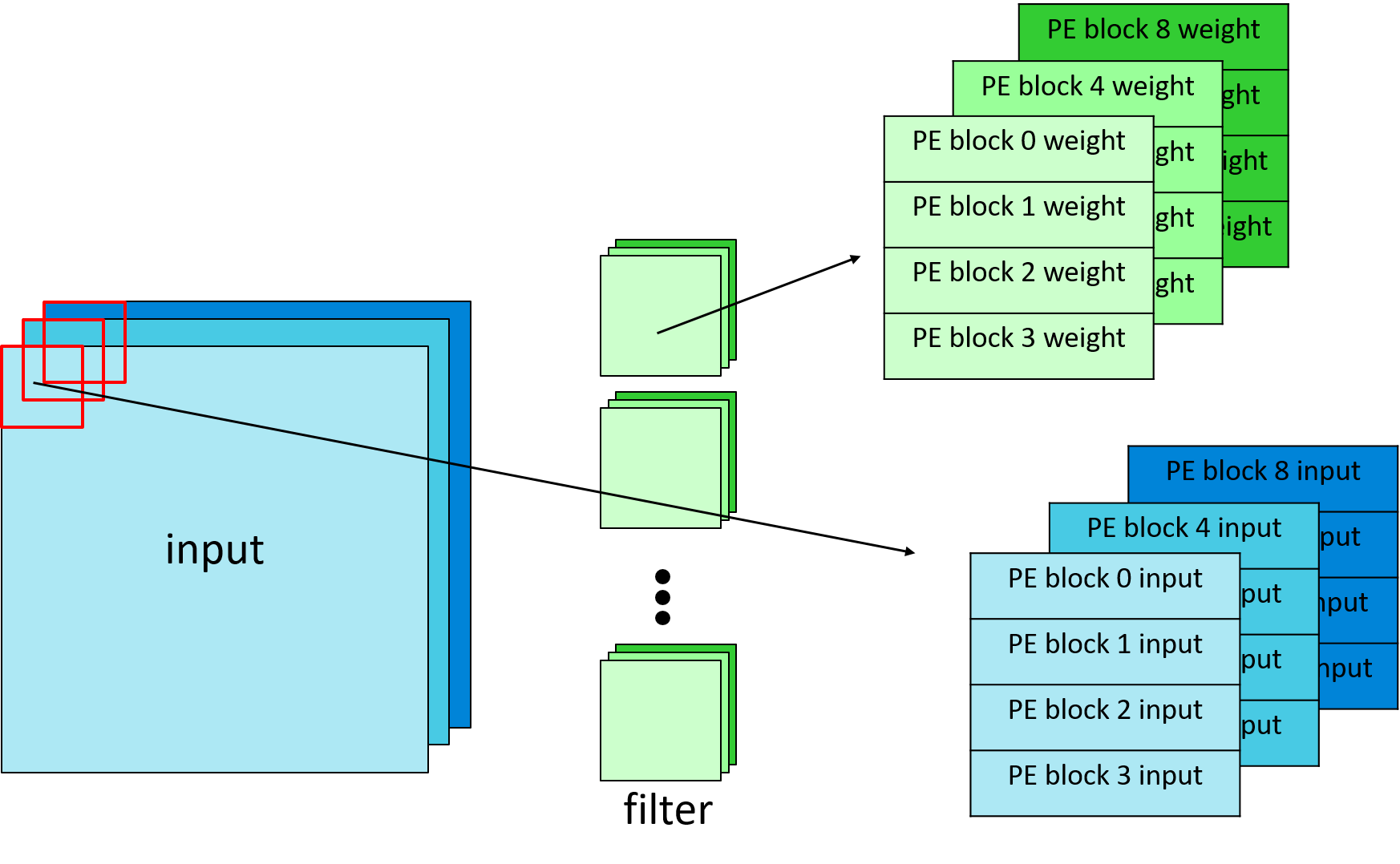}}
\subfigure[]{
\label{conv detailed}
\includegraphics[width=0.23\textwidth]{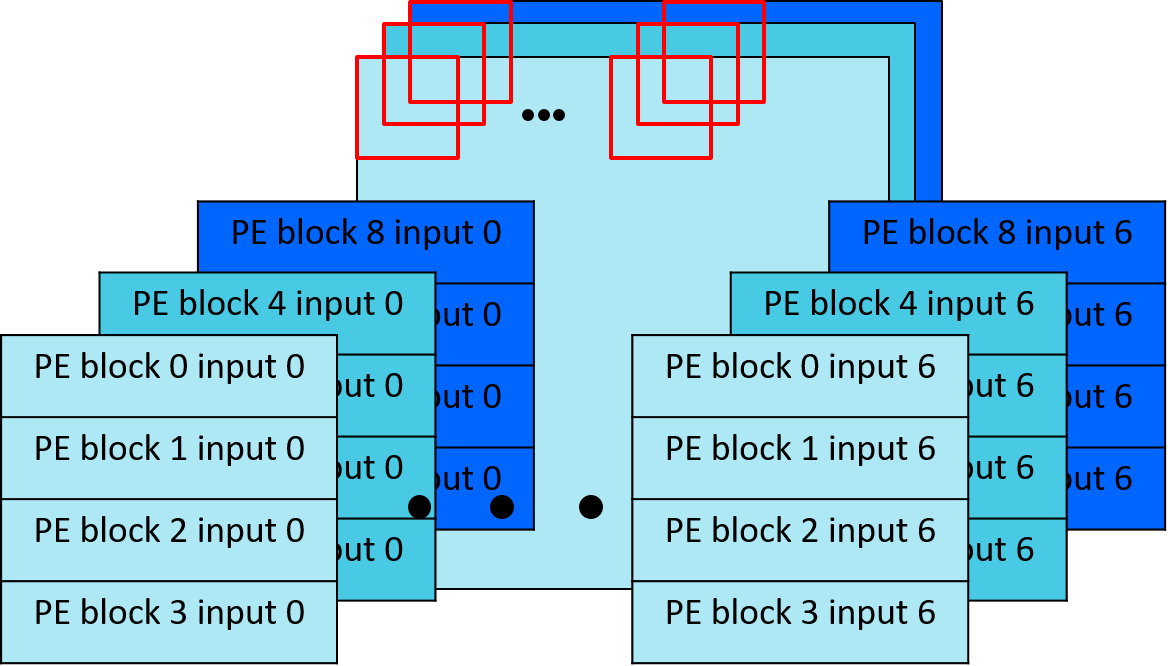}}
\caption{(a)Input and weight mapping to PE blocks for one output convolution channel. (b) Detailed mapping for input of PE blocks}
\end{figure}

The only convolution in vision transformers is at the first layer, which uses 4×4 convolution with stride 4 to reduce the input size into 1/4 of the original. The size of each kernel is 4×4×3, which can be perfectly placed into each PE weight block, as shown in Fig.~\ref{PE for convolution}. The input is RGB images. Each input channel and its corresponding weight will be placed on four PE blocks, respectively. During each computing cycle as in Fig.~\ref{conv detailed}, all 7 input rows will be activated at the same time to process all input to generate 7 outputs in a cycle. The computing order of convolution will process the first output channel and then the second one until the final 96th channel. For the  224×224×3 input size in ImageNet, our design would calculate a 28×4×3 input in a cycle and take 448 cycles to finish one output channel.

\subsection{Fully-Connected Layer}
By row-wise scheduling, different sizes of fully connected layers can become a dot product in a single row. However, since the number of input channels are multiple of 96 and exceeds the hardware capability in one cycle, its output will be accumulated in the accumulator for the final output.

\begin{figure}[htb]
\centering
\subfigure[cycle 1]{
\label{linear1}
\includegraphics[width=0.15\textwidth]{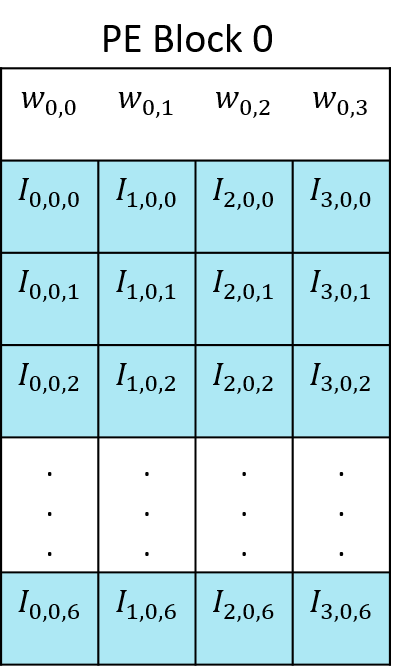}}
\subfigure[cycle 2]{
\label{linear2}
\includegraphics[width=0.15\textwidth]{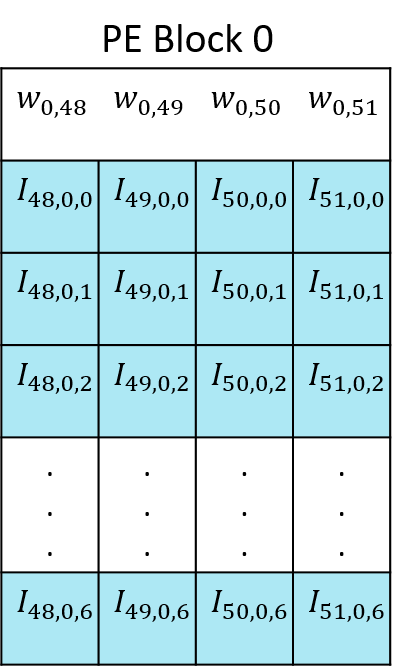}}
\caption{Detailed input and weight mapping of 7 inputs for a fully-connected layer with 96 channels at (a) the first cycle and (b) the second cycle. }
\end{figure}

Fig.~\ref{linear1} and Fig.~\ref{linear2} show the detailed calculation for 96 channels. In the first cycle, the 1st to 48th channels of input and filters would be placed at the corresponding position from PE block 0 to 11. The second cycle will process the rest of channels. Besides, with 7 input rows in a PE block, we would finish 7 outputs in every 2 cycles for 96 channels of input. Then, the computing order of the fully connected layer will process the first output channel and then the second one until the final output channel.

\subsection{Multi-Head Attention Layer}
With the row-wise scheduling and the window multi-head self-attention(W-MSA) concept as in Swin-Transformer, we can make the computation of multi-head attention layers regular and simple. In this case, for the multiplication of Q and K as in (\ref{eq self attention}), Q could be regarded as the weight and K as the input for PE blocks. In this case, since the size of the matrix is smaller than other layer types, it will only use 8 PE blocks. Thus, for Q matrix mapping, four columns of Q matrix will be allocated to one PE block and processed in a row-by-row manner. Each row will take 7 cycles. For K$^T$ matrix mapping,  they are allocated to a group with 7 input rows × 8 PE blocks for computation, and processed in a group-by-group manner. During this process, since only 8 of the 12 PE blocks, the hardware utilization will be lower. However, the flops of the multi-head attention layer would take no more than 3\% in whole model, and such hardware utilization impact would impact no more than 1\% of total cycles.

\section{Experimental Results}
This design has been implemented with TSMC 40nm CMOS technology and needs 262K NAND gate count and 149KB SRAM buffer to execute vision transformer models, especially optimized for Swin Transformer. The precision of weights and activations are both 8 bits.  For area, SRAM occupies 93.2\% of area, and PE, SRAM controller, and accumulator occupy 4.6\%, 1.5\%, 0.6\% of area respectively. The peak performance is about 403.2GOPS at 600MHz clock frequency. The average time to process an 224$\times$224 image with Swin-T is about 22.4 ms, equivalent to 44.5 images/s throughput. Theoretically, overall hardware utilization could be as high as 99\% or higher. 

Comparisons with other designs are not easy due to different target models. For reference, we list other CNN hardware accelerators in Table~\ref{table:hardware compare}. Our peak throughput is higher than others due to more PE numbers and higher clock frequency. However, our area cost is smaller than others due to its simple PE structure.

Table~\ref{table:fpga compare} shows the throughput comparison to execute Swin-T with different processors and designs. GPU is a NVIDIA GeForce RTX 2080 Ti, which has 4352 CUDA Cores and a total of about 9.8MB of cache memory. Compared to GPU, our throughput is 1.1$\times$ than that of GPU, respectively. This higher throughput is achieved by our high hardware utilization. Another design is based on FPGA.  Vis-Top~\cite{hu2021vis} is 1.9$\times$ faster than us, but little information has been disclosed in this paper. This higher throughput is possible due to larger PEs based on its DSP slices utilized.

\begin{table}[]
\centering
\caption{Comparison to other CNN accelerators}
\label{table:hardware compare}
\begin{tabular}{|l|c|c|c|}
\hline
                      & Ours   & IECA~\cite{huang2021ieca}  & Eyeriss v2~\cite{chen2019eyeriss} \\ \hline
Technology(nm)        & 40     & 55    & 65         \\ \hline
Model Types           & Trans. & CNN   & CNN        \\ \hline
PE Number             & 336    & 168   & 92         \\ \hline
Clock Rate(MHz)       & 600    & 250   & 200        \\ \hline
Peak Throughput(GOPS) & 403.2  & 84    & 153.6      \\ \hline
Area (KGE)            & 186    & 344.7 & 2695       \\ \hline
SRAM(KB)              & 149    & 109   & 192        \\ \hline
\end{tabular}
\end{table}

\begin{table}[]
\centering
\caption{Comparison to other processors and designs.}
\label{table:fpga compare}
\begin{tabular}{|l|c|c|}
\hline
                     & GPU                       & Ours (ASIC) \\ \hline
Throughput (image/s) & 41.5                      & 44.5        \\ \hline
Relative Speedup     & 1                         & 1.07        \\ \hline
Cache Memory(KB)     & 9852                      & 149         \\ \hline
Throughput per MACs  & 0.0095                    & 0.1324      \\ \hline
\end{tabular}
\end{table}

\section{Conclusion}
This paper proposes an efficient design for vision transformers that is optimized for encoding layer operations instead of CNN. This design can efficiently execute different layer operations by row-wise scheduling for high hardware utilization. Our design retains high throughput while maintaining area efficiency. The final implementation with TSMC 40nm CMOS process needs 262K NAND gate counts and 149KB SRAM buffer and achieves 403.2 GOPS peak throughput when running at 600MHz clock frequency. 
\bibliographystyle{IEEEtran}
\bibliography{bibliography.bib}

\end{document}